\documentclass[openacc]{rstransa}
\usepackage{xcolor}

\titlehead{Research}

\newcommand{\ML}[2]{\st{#1}{\bf \textcolor{olive}{#2}}}

\usepackage[pagewise]{lineno}
\usepackage[normalem]{ulem}

\begin{document}

\title{\bf Future Prospects for Partially Ionised Solar Plasmas: the Prominence Case.}
\author{
S. Parenti$^{1}$, M. Luna$^{2}$, J. L. Ballester$^{2}$}

\address{$^{1}$Universit\'e Paris-Saclay, CNRS, Institut d’Astrophysique Spatiale, 91405, Orsay, France. \\$^{2}$Departament de F\'\i sica and  Institute of Applied Computing \& Community Code (IAC$^3$). Universitat de les Illes Balears, 07122 Palma de Mallorca (Spain). \\
 }
\subject{XXX}


\corres{S. Parenti\\
\email{susanna.parenti@universite-paris-saclay.fr}}


\begin{abstract}
Partially ionised plasmas (PIP) constitute an essential ingredient of our plasma universe. Historically, the physical effects associated with partial ionisation were considered in astrophysical topics such as the interstellar medium, molecular clouds, accretion disks and, later on, in solar physics.  PIP can be found in layers of the Sun's atmosphere as well as in solar structures embedded within it.  As a consequence, the dynamical behaviour of these layers and structures is influenced by the different physical effects introduced by partial ionisation. 

Here,  rather than considering an exhaustive discussion of partially ionised effects in the different layers and structures of the solar atmosphere, we focus on
solar prominences. The reason is that they represent a paradigmatic case of a partially ionised solar plasma,  confined and insulated by the magnetic field,   constituting an ideal environment to study the effects induced by partial ionisation.  We present the current knowledge about the effects of partial ionisation in the global stability,  mass cycle and dynamics of solar prominences.  We revise the identified observational signatures of partial ionisation in prominences. We conclude with prospects  for PIP research in prominences,  proposing the path for advancing in the prominence modelling and theory and using new and upcoming instrumentation.

\end{abstract}


\maketitle

\section*{Introduction}
\label{intro}
The term ``plasma universe'' was introduced by \cite{alfven86}
to highlight the important role played by plasmas across the universe. The study of plasmas beyond Earth's atmosphere is called ``plasma astrophysics'',  and partially ionised plasmas (PIP) constitute an essential ingredient of many astrophysical plasmas such as the Sun, the heliosphere, magnetospheres of the Earth and planets, the interstellar medium, molecular clouds, accretion disks, exoplanet atmospheres, stars and astrospheres, cometary tails, exoplanetary ionospheres, etc.  A recent summary on PIP dynamics in astrophysics can be found in \cite{ballester18} and references therein.

In these environments, the ionisation level varies from almost no ionisation in cold regions to fully ionised in hot regions,  and PIP introduce physical effects which are not considered in fully ionised plasmas and which are crucial to fully understand the behaviour of astrophysical plasmas in different environments.   Regarding solar physics,  in recent years the study of PIP has become a hot topic because solar structures such as spicules,  prominences,  as well as layers of the solar atmosphere (photosphere, chromosphere, transition region) are made of PIP,  while,  at the same time,  other dynamical processes,  such as waves and instabilities,  are also influenced by partial ionisation effects.  In general, to study PIP
which consists of electrons, ions, and neutral atoms, it is assumed that each species has a Maxwellian velocity distribution, therefore they can be described as separate fluids. Then,  the equations in the three-fluid description can be written and from them the transition to two-fluid and single-fluid approaches can be made.  In the single-fluid approximation all the plasma species are considered strongly coupled,  then  this approach allows the formulation of governing single-fluid equations that describe the plasma as a whole,  using global or averaged quantities,  but interactions between different species appear in the form of non-ideal terms such as ambipolar diffusion. The single-fluid approach is valid for time scales that are longer than the ion--neutral collision time but it fails for shorter time scales.  Then,  the two-fluid approximation,  which means that ion--electron and neutral gases are treated as separate fluids, and which is valid for time scales longer than the ion--electron collision time,  should be considered  \cite{zaqa11}.  In this case,  interactions between species involve different processes such as momentum transfer,  particle collisions, ionisation, recombination,  charge exchange,  etc... \cite{ballester18}.  Summarising,  it is of great theoretical and observational importance to fully understand the differences introduced by the use of multi-fluid approaches versus the single-fluid approximation,  which could lead to completely new physics whose observational signatures \cite{gonzalez_manrique_two_2024} could be detected by present and forthcoming observational facilities.

Regarding solar atmospheric layers,  the photosphere, chromosphere, and the transition region are partially ionised layers of the solar atmosphere. Therefore,  the interaction between ions and neutrals strongly influences the dynamical behaviour of these layers showing important differences with respect to the fully ionised case.  While recent numerical simulations of the chromospheric plasma involve a multi-fluid approach with a finite plasma-beta including additional physical effects of partial ionisation and radiative transfer in NLTE  \cite{ballai19, soler12b,  soler19},  it is still possible to treat the chromospheric plasma as a single MHD fluid \cite{sykora15} including the most important partial ionisation effects. However, the validity of this approximation in the chromosphere is not general since it depends on the ion--neutral and neutral--ion collisional scales relative to hydrodynamic scales. 

 In order to understand the effects produced by ambipolar diffusion in the chromosphere,  in a review by \cite{sykora15} 2.5D radiative MHD simulations have been used to conclude that these effects are: dissipation of magnetic energy,  increase of the minimum temperature in this layer, and heating \cite{sri21} and expansion of the upper chromosphere. The knowledge and understanding of the physics of the chromosphere is of crucial importance in order to establish a meaningful, dynamic background useful to perform detailed studies related to the propagation and dissipation of MHD waves \cite{soler15a,  soler19,  melis21}.

Among other solar structures, solar prominences are chromospheric structures embedded in the solar corona.  Because of their low temperature,  prominence plasma is partially ionised,  although the exact degree of ionisation is still unknown and,  probably,  varies over a wide range with the ratio of electron-to-neutral hydrogen density roughly between 0.1 and 10 \cite{patso02}.  Partial ionisation contributes the presence of neutrals and electrons in addition to ions,  thus collisions between the different
species are possible and the effects on the prominence equilibrium and dynamics should be considered.  Furthermore,  dynamical features observed in prominences have been interpreted and modelled in terms of different instabilities,  or in terms of MHD waves \cite{arregui18}.  In most cases,  these interpretations have considered the plasma to be fully ionised,  however  partial-ionisation effects produce modifications of the instability thresholds and growth rates \cite{soler12a,  soler12b,  diaz12,  khomenko14,  diaz14,  martinez15,  ballai17,  ruderman18,  soler22} and influence the damping rates and energy dissipation of MHD waves  \cite{soler13a,  soler13b,  soler13c,  soler15,  martinez16,  martinez17,  martinez18,  ballester18} (see Soler's review article on MHD waves).  Also,  in some cases, multi-fluid physics needs to be incorporated into existing models of oscillatory phenomena and instabilities in order to fully understand the prominence dynamical behaviour.  However,  in spite of these advances,  when partial ionisation is taken into account, several key problems related to prominence physics still remain unsolved (see next sections).

In summary,  the role played by partial ionisation in the stability,  energetics,  and dynamics of solar structures as well as in solar atmospheric layers is of great interest,  therefore,  research on these topics,  involving in a consistent manner partial-ionisation effects,  has already started and,  in the near future,  further developments are anticipated.  
As we have stated in the previous paragraph,  solar prominences represent a paradigmatic case of a partially ionised solar plasma,  
One of the advantages of studying PIP in prominences is that they stand out against the hotter and less dense corona, allowing  better isolation of their emission from the  contribution of the rest along the line of sight, and thus enabling a better interpretation of the observations to be compared to theoretical predictions. Furthermore, with respect to the complex chromosphere, modelling the PIP effects in prominences can be done with a more realistic approach.
This is particularly true when addressing the prominence--corona transition layer, which is the boundary between the core of the feature filled by neutrals (or partially ionised plasma), and the completely ionised corona. 
Therefore,  rather than reviewing the effects produced by partial ionisation in all of the solar atmospheric layers above mentioned,  in
the following sections we will focus on solar prominences suggesting some tentative theoretical and observational approaches which,  in the near future, could help to understand the physics of partially ionised solar prominences.

\vspace{1cm}

\section*{Theoretical modelling PIP in prominences: current insights}
\label{model}
 Solar prominences (also called filaments when seen on the solar disk) are clouds of cold and dense plasma in the hot and tenuous solar corona. 
Although much progress has been made in recent years, there are still many open questions in the modelling of solar prominences. 

The PIP physics  in prominences plays a key role, such as in their global stability, mass cycle, and dynamics.
%
%
Prominences are hosted in an elongated magnetic field   structure called a filament channel. 
These magnetic structures appear and evolve as a result of mechanisms associated with the dynamics of the photosphere and chromosphere. This channel may exist in the absence of prominences, suggesting that their evolution may be, in principle, independent \cite{mackay_physics_2010}.
There is consensus that it is the magnetic field that supports prominence mass against gravity and insulates it from the surrounding hot corona.
The cold plasma is located in the upward concavities of the magnetic field lines: the dips. However, the ability of the field to support PIP has not been studied in sufficient detail and only a few studies have considered this aspect \cite{gilbert_neutral_2002,terradas_support_2015}. The neutral plasma does not interact with the magnetic field so it would be expected to neither fall nor ever reach these heights. 

The ion--neutral coupling is key to preventing the rapid draining of the neutral plasma from the prominence body.
One of the first papers to consider this problem was \cite{mercier_downward_1977}. The authors -- somewhat surprisingly -- found that, due to frictional coupling between the neutral and ionised hydrogen, the plasma would fall very slowly.
Later, \cite{gilbert_neutral_2002} considered a plasma composed of ionised and neutral hydrogen and helium. They also included the charge-exchange coupling between the ionised and neutral hydrogen atoms. They found that neutral hydrogen was strongly tied to the ions, but neutral helium could be drained out of the prominence relatively quickly. They calculated that helium is completely drained from the prominence in one day, in contrast to the 22 days for hydrogen.
From this work, it seems that 
neutrals heavier than hydrogen escape from the prominence more easily. 
An observational study by \cite{gilbert_observational_2007} found a deficit of neutral helium in the upper part of the prominences and a surplus in the lower part indicating that helium drainage is responsible for this element's distribution.
These theoretical works were based on simplified models that did not include the complexity of the prominences.
In a more elaborated model by \cite{terradas_support_2015}, two-dimensional numerical simulations were performed in the two-fluid approximation for a plasma composed only of hydrogen. They found that neutral hydrogen is strongly coupled to its ions mostly by charge exchange confirming the results from \cite{gilbert_neutral_2002}. 
In these works, the authors considered monolithic prominences in which the neutrals drain from top to bottom throughout the whole structure. In contrast, observations indicate that prominences are composed of thin threads  150\,--\,450 km thick \cite{lin_thin_2005,lin_filament_2008}. 
These strands are horizontally piled, forming the body of the prominence.
%
%
%
 With this consideration, the draining times should be reduced by at least one order of magnitude. The neutral plasma would leave the threads within an hour to a few days.%
This topic is very relevant for understanding the stability of solar prominences and may also be relevant for the mass cycle in these structures. 
 In-depth studies are needed to consider more realistic configurations, including additional species in the plasma such as helium, as well as incorporating extra terms in the equations, such as photoionisation and recombination, to account for variable ionisation fractions.

 In the solar corona, thermal conductivity is often considered to be essentially parallel to the magnetic field. However, this is only true for fully ionised plasma and above a certain temperature \cite{spitzer1962}. 

In partially ionised, cold plasmas, an isotropic conduction term appears, allowing conduction perpendicular to the magnetic field.
In this way, the  prominence cold threads in contact with the surrounding corona will be affected.
Studies incorporating transverse conduction due to the presence of neutrals would be necessary to understand the thermodynamic equilibrium between the corona and prominences. 
 Only a few works \cite{soler12a,popescu_braileanu_two-fluid_2021,popescu_braileanu_two-fluid_2021b,popescu_braileanu_magnetic_2023} have studied different instabilities in prominences  analytically and numerically.
The authors show that the presence of neutrals increases the growth rate of thermal instabilities and favors the formation of small scale structures during the development of RTIs.
It is clear from  these works that it is necessary to go further, into the non-linear regime and non-uniform plasmas with more complex models.
 For instance, significant progress has been achieved in this direction with the implementation  of the full conductivity tensor in the MANCHA3D code \cite{navarro_modeling_2022,modestov_mancha3d_2023} but only for fully ionised plasmas. In its two-fluid version, MANCHA3D-2F, only neutral conductivity is included \cite{popescu_braileanu_two-fluid_2021,popescu_braileanu_two-fluid_2021b,popescu_braileanu_magnetic_2023}. It is anticipated that both conductivities will be incorporated in a future version of the code.
The prominence mass cycle encompasses the set of mechanisms leading to the formation of the cold prominence plasma, as well as the processes that contribute to its eventual disappearance.
It is generally believed that the prominence mass has a chromospheric origin because there is not enough plasma in the corona to form them \cite{pikelner_origin_1971,saito_arch_1973}. 
Several models have been proposed to explain how chromospheric plasma comes to form the prominences: injection, levitation, or evaporation--condensation mechanisms. There may be no single mechanism that universally explains the formation of all prominences. However, the evaporation--condensation mechanism is the most plausible one to explain the large prominences that can persist for weeks or even months \cite{luna12,karpen_plasma_2015}.
According to this model, coronal heating evaporates the chromospheric plasma that flows along the field lines of the filament channel. 
This plasma cools down and condenses to produce the threads, which are the building blocks of the prominences.
 Condensations form due to thermal instability: an imbalance between radiative losses, thermal conduction, and coronal heating. The temperature decreases and density increases until a new equilibrium is established.

This process has been simulated many times, always using fully ionised plasmas \cite{antiochos99,karpen_condensation_2008,luna12,xia14}. 
 However, partial ionisation can play a crucial role as the ionised million-degree plasma cools down to about 8000\,K. 
In this cooling process, the fully ionised plasma becomes partially ionised, predominantly neutral.
Electrons recombine with ions to form neutral atoms.

%
The recombination releases  additional energy modifying the condensation process.
 During condensation, the gas pressure drops locally producing the accretion of the surrounding plasma increasing the density.
The recombination reduces the electron gas pressure in the core of the threads. This could accelerate the condensation process as it favors 
 mass accretion.
Thus, we see that it is relevant to consider the ionisation--recombination energy to understand the formation of the cold mass of the prominences. 
%

Evaporation--condensation models succeed in explaining the formation process and the mass of the prominences \cite{luna12,xia_simulations_2012}. 
However, they predict a continuous growth of the prominence. 
It is unclear what happens to the threads after their formation, but these models fail to explain the complete mass cycle.
Perhaps a dynamical equilibrium is established, where evaporation injects hot plasma into the prominence and by some process, the prominence loses mass.
This mass loss may be caused by the plasma falling from the prominence into the chromosphere, i.e. the drainage, which has already been observed
(see, e.g., \cite{engvold_small_1981,kubota_vertical_1986,berger_hinode_2008,liu_first_2012}). However, it is unclear whether it is a common phenomenon or limited to certain types of prominences.
Some authors have interpreted these  vertical draining motions in terms of different instabilities, e.g. Rayleigh--Taylor, Kelvin--Helmholtz, and thermal instability \cite{ryutova10,  hillier11,  hillier12,  hillier18a,  hillier18b,  berger17}. These authors considered these instabilities in fully ionised plasmas. However, \cite{khomenko14} studied the Rayleigh--Taylor instability in PIP and the effect of ambipolar diffusion. More recently, \cite{popescu_braileanu_two-fluid_2021,Popescu21,popescu_braileanu_magnetic_2023} studied PIP in the two-fluid approximation. The authors showed that 
 charged and neutrals decouple 
{\bf having} a significant impact on the instability  plasma dynamics.
 However, they found that ionisation--recombination had a limited impact on the dynamics. 
As the authors point out, they do not include optically thin radiative losses which, as shown by \cite{ni_magnetic_2018}, can increase the impact of ionisation--recombination on the plasma dynamics.

In this way, neutrals may accelerate destabilisation and be relevant in the prominence-drainage process.
Thus the process of condensation together with drainage due to the neutrals' leakage  can be a cycle that can maintain stable prominences for weeks or even months.
%

 Alternative theoretical investigations offer a novel perspectives on PIP physics; 
 for example the study by \cite{miloch_plasma_2012} based on particle-in-cell (PIC) 3D simulations. The authors studied the interface between the cool and dense plasma of a prominence and its tenuous and hot surrounding corona.
The magnetic field was considered parallel to the interface and the ion--neutral coupling effects were incorporated. 
The results show that anomalous Bohm diffusion occurs at the interface yielding the mixing of the two plasmas. 
 In this process, ion--neutral collisions are of little importance.
 The anomalous Bohm diffusion, characterised by the movement of particles across the field, has been documented in numerous plasma experiments \cite{chen_plasmafusionbook_2016}.
 However, these simulations have limitations. The authors consider a very small numerical cubic box with a side of less than 10 cm and a simulation time of microseconds. The transition zone is expected to be several orders of magnitude larger than the size of that domain \cite{chiuderi_energy_1991}. With such short evolution times, it is possible that a stable situation has not been reached. Perhaps, if further evolution is allowed, the transition zone would continue to grow until the domain boundaries are reached.
In the future, it will be interesting to perform simulations of this style but with more complexity to understand PIPs in solar prominences and their transition zones.
%

 \section*{Observational evidence of PIP in prominences: current insights }
The proposed effort for the modelling of prominences needs to be accompanied by the improvement of our observational tools to diagnose and quantify the presence of PIP, and to validate its physical influence on the features as predicted by the modelling. 

Due to their favorable observation at the limb against a much diluted and highly ionised corona, prominences are indeed good candidates to validate the amplitude of the impact of PIP physical effects predicted by multi-fluid plasma models and simulations. This advantage is also given by the fact that PIP effects are amplified by the dynamic nature of prominences and are predicted to be visible at the interface between the corona and the cool prominence core. 

In fact, prominences show signatures of dynamics motions over a wide range of spatial and temporal macroscopic scales. These include mass flows, waves propagation and/or damping over coherent large scales, or smaller bunches of threads and streams in opposite directions (up and down streams for prominence, counter-streaming for filaments, see also \cite{parenti_solar_2014} for a review). Instabilities, waves, and turbulent motions are also present over smaller scales, where the ion--neutral decoupling can be important. 

The dynamics associated with multi-fluid plasma physics have been predicted to play a role over very small scales, often below the observational limits of our instrumentation ($< 1$ mins, ~ 10 Km, e.g. \cite{Khomenko12, gonzalez_manrique_two_2024}). However, as illustrated previously, it may also have an impact on the stability and lifetime cycle of the whole prominence.  
 One of the manifestations of the ion--neutral partial or total decoupling is a different velocity amplitude of these observed species, also called velocity drift. This has been predicted to vary depending on the physical process considered and on the property of the layers involved. The amplitudes are  from about a few hundreds m/s to a few Km/s e.g. \cite{khomenko14, popescu_braileanu_two-fluid_2021}. The lower value is, in some cases, at the limit of the capability of the available instruments/observation conditions.
 Inferring the Doppler drift  is also difficult due to the limitation of our observations: the emissions from ions and neutrals may originate from different regions along the line of sight, providing a misleading interpretation of the Doppler drift.
An example is given by the presence of optically thick plasma along the line of sight, as in the case of prominences observation. 
Under this circumstance it is possible to obtain different values of ion--neutral drift, both using lines from different ions of the same element, and lines from different species.
The solution to this problem is generally given by the selection of that fraction of the available data that relates to (and are deduced to be) optically thin plasma regions. The integration of the emission along the whole line of sight would, in principle, minimize the effect due to the possible presence of still optically thick sub-regions. 
Aware of these limitations, a few investigations have been carried out providing some observational evidence and their possible interpretation. 
Such examples, however, reveal the difficulty of the task: the velocity excess of the ions has been measured in extended areas of quiescent (e.g. \cite{wiehr19, zapior22}) and dynamically evolving prominences \cite{wiehr21} while \cite{khomenko16} detected it only occasionally, in the transient higher velocity localized areas. While it appears repeatedly that a more dynamic environment is associated with detectable ion--neutral velocity drifts, the persistent presence of the drift in a quiescent prominence should  somehow be explained. The typical timescale of the decoupling is much shorter than the lifetime of a quiescent prominence. The strength and the duration of the drifts are influenced by the magnetic field amplitude, as it enters the strength of the Lorentz force on ions. Ion--neutral collisions depend on the density: a low density reduces the ion--neutral interaction and their coupling. A more systematic and quantitative observational analysis of the relation between ion--neutral decoupling and these parameters is still lacking.

 Similarly, spectral line width measurements are a way to possibly detect the amplitude of the ion--neutral decoupling within unresolved motions. 
 The width of a spectral line in prominences is interpreted as due to two possible dominating effects: a thermal component (kinetic temperature), given by the temperature of the observed species, and a non-thermal component, possibly due to either turbulence, waves, or general unresolved motions.
To derive the Doppler width as a diagnosis of ion--neutral decoupling, we need the measure of several spectral profiles from neutrals and low ionisation state species under optically thin condition, with the further approximation that the lines are formed within the same volume. 

The literature provides results where the line width from ions is larger than that from the neutrals. The interpretation of this result depends on the assumed hypotheses. Assuming the non-thermal component to be larger for ions than for neutrals,  \cite{gonzalez_manrique_two_2024} measured line widths consistent with a unique kinetic temperature. Other authors,  instead, found different thermal components for the two species under the assumption of a zero non-thermal motion (e.g. \cite{ramelli12, stellmacher15, stellmacher17, gonzalez_manrique_two_2024}). Knowing that  prominences are made up of dynamic fine, probably unresolved, structures, we expect to observe  the result of the superposition motions, both within the instrument's resolution and,  under optically thin plasma assumption, along the line of sight. Thus, the first interpretation is possibly the most probable. 

The important aspect of interpreting these observational signatures as the imprint of one or more of the candidate physical processes acting in PIP is still not straightforward. For this, we need the support of the modelling that could identify and quantify the unique imprint for each process.
 One example is given by \cite{Popescu21}, who investigated the sensitivity of the Rayleigh--Taylor instability (RTI) to the ion--neutral collisions and their consequent decoupling, at the interface between the corona and the prominence. They found a Doppler shift of about an order of magnitude larger than that predicted in the absence of the RTI \cite{gilbert_neutral_2002, terradas_support_2015}: the interfaces where a RTI can develop are candidate sites to better measure the ion--neutral decoupling.

We already mentioned that the magnetic field amplitude and direction affect the Doppler shift. Measures of the magnetic field in prominences are quite limited and have important uncertainties.  For instance, the line of sight, the different height in the atmosphere where the spectral lines are formed, and the 180\textdegree~ ambiguity, can all affect the polarimetric diagnostics \cite{valori_2023}. 
Another limitation up to now is given by the sensitivity of  polarimetric instruments, which should be high enough to be able to measure the weak magnetic field in prominences (1\,--\,few tens of Gauss for quiescent prominences, \cite{parenti_solar_2014, mackay_physics_2010}) with a relatively high cadence, as done for measuring the temporal variation of the plasma parameters. 
The existing literature reports active-region prominences to have a stronger field than the quiescent ones \cite{parenti_solar_2014}, but it is still not clear if such a difference is detectable, for instance, in the Doppler velocity shift.
This is a further reason for increasing the effort in providing systematic measures of magnetic field and Doppler shift in prominences. 

In summary,  to better understand the PIP physics in prominences, we need to face the problem of providing  co-temporal, high spatial and temporal resolution measures of Doppler width, shift of neutrals and ions,  together with the vector magnetic field.  We need  to have instruments sensitive enough to measure the Doppler shift of fractions of a km/s, to have an independent measure of the kinetic temperature and non-thermal velocities (using emission from multiple ions and neutrals at similar temperature). 

Recalling that the PIP effects arise over small spatial scales, we need to ensure to measure the plasma and magnetic field properties of the same volume. In addition to the line-of-sight problem mentioned before, we require an excellent co-registration of the various instruments adopted for the measures. Only in this way, can we quantify the effect of the magnetic field on the ion--neutral decoupling. 
These should be all reference conditions to validate the various models that describe the ion--neutral decoupling and its amplitude under different physical conditions. 

 The other aspect that needs attention in order to have a better impact of our observational data is the further development of diagnostics tools for optically thick plasma and radiative-transfer treatment for prominence applications. It is a complex problem, and the application of forward modelling is one way that the diagnostics can be performed over several spectral lines (see \cite{Labrosse2010} for a review for non-LTE modelling in prominences).
These generally use mult-ilevel models for some or all of the H, He, C, and Mg species (see also next section). 
Other, more general codes exist, which perform spectral-line inversion for photospheric and  chromospheric plasma environments, to recover magnetic field and plasma properties. However, only a small portion of them have been used for prominences investigations \ML{(e.g. STIC, NICOLE)}{(e.g. STIC \cite{rodriguez_stic_2019}, NICOLE \cite{socas-navarro_open-source_2015}, HAZEL \cite{asensio_ramos_advanced_2008})}: they represent an existing potential for prominences application.

In the following section, we  conclude by presenting and discussing some key theoretical and observational novelties for future research of PIP in prominences.

\section*{Prospects for future research}

 Quantifying the ionisation degree of the plasma in prominences is important,  as it provides the strength for the PIP effects; it enters for instance in the derivation of the mass, and it affects the plasma motion properties along and across the magnetic field. We have seen in the previous paragraph that PIP effects generally act locally (but with consequences also on the whole prominence), which requires high resolution observations to be detected. We then need models that also increase their complexity to become more realistic in their mimicking of the high resolution properties of these features. 
 The ionisation state, in principle, can be computed taking into account all of the ionisation and recombination processes and, then, to determine the time-dependent values of the ionised and neutral fractions.  \cite{heinzel15} took a different approach to determine the ionisation degree in several prominence slabs. In particular, these authors considered a prominence plasma composed of hydrogen and fully neutral helium whose abundance was 10\,\%. Then, using the H$\alpha$ spectral line and one-dimensional non-LTE radiative transfer models \cite{heinzel14} they provided tables for the ionisation degree for diﬀerent temperatures and pressures at different heights inside the prominence slab. However, it is highly probable that theoretical calculations do not reflect the true behaviour of the ionisation degree because taking into account the very inhomogeneous and dynamic prominence plasma, depending on space and time, makes its theoretical and observational determination much more difficult. Furthermore, it is worth pointing out that the ionisation structure inside prominences influences processes such as thermal conduction, ambipolar diffusion, wave damping, coupling of prominence plasma to magnetic field, etc... relevant in partially ionised plasmas.

 Another important issue in prominence research is the coupling of radiation to magnetohydrodynamics, which has been explored in the case of fully ionised plasmas. For instance, \cite{heasley76, anzer99} considered a Kippenhahn--Schl\"uter model and the magnetohydrostatic equilibrium was solved coupled to NLTE radiative transfer for multi-level hydrogen and helium atoms. This model was generalised to 2D geometry, coupling magnetohydrostatic Kippenhahn--Schl\"uter equilibrium to 2D NLTE radiative transfer in a multi-level hydrogen \cite{heinzel01}. Over the past years, modelling has been improved by considering 2D or 3D NLTE simulations together with arbitrary magnetic fields filled with prominence plasma \cite{gunar13}. However, and to the best of our knowledge, no attempt has been made to couple radiative transfer with magnetohydrodynamics in the case of partially ionised prominences.

 In addition to these examples of the open issues in prominences modelling,  numerical code development is progressing rapidly, making it possible to include most PIP effects in the near future. These studies must address the complete structure of the solar prominences, considering both the exchange of mass and energy with the chromosphere and the corona and the fact that the cold plasma is composed of thin threads, detectable at the limit of our instrumentation. This complexity leads to a wide range of spatial scales, requiring numerical simulations across large numerical domains with high spatial resolution. 

 Today, several codes allow numerical simulations of prominences such as MPI-AMRVAC \cite{keppens_mpi-amrvac_2023}, MANCHA3D \cite{modestov_mancha3d_2023},  Bifrost \cite{gudiksen_stellar_2011}, or PIP code \cite{hillier_formation_2016} among others. MPI-AMRVAC has implemented fully ionised plasma thermal conduction, optically thin and adaptive-mesh radiation (AMR), which allows numerical experiments on prominence formation by TNE. MANCHA3D or Bifrost are those that include more PIP terms in the single-fluid approximation. Both codes extend from sub-photospheric layers to the corona and also include magnetoconvection. They could be used for conducting self-consistent numerical experiments of prominence formation and evolution. Such codes have sacrificed AMR as it is not needed in the photosphere or chromosphere. Running simulations with these codes can be computationally expensive. However, with the development of high-performance computing facilities, such numerical experiments may become feasible.

 In the upcoming decade, there will be an increase of available instruments for solar observations. More frequent multi-observatory, co-temporal data will be available to allow for the derivation of multiple ions, neutrals, and magnetic field parameters. To take advantage of these data, however, we need to increase the effort to further develop the diagnostic tools for multiplr point of view observations. There is a great potential for a more accurate constraints and limiting the free parameters of the modelling and simulations of PIP.

Solar Orbiter, for instance, is the latest ESA solar mission (launched in 2020 \cite{muller20}) and carries, among others, the EUI suite \cite{rochus2020} with a high spatial and temporal EUV imager centered at 174~ \AA~~(Fe\,{\sc x}), the HRIEUV. During the Solar Orbiter perihelion at around 0.29 AU, this instrument records the dynamics of the solar atmosphere at a maximum of 3s cadence with a spatial resolution of about 200 km. This waveband has already shown great capability of resolving prominence fine structure  \cite{berghmans2023}.  This can be used to study the dynamic nature of threads and inferring, using seismology diagnostics, magnetic field strength at such small scales\cite{arregui_2011}. It is also potentially possible to study the effect of PIP on wave damping in threads, using coordinated observations with other ground-based observatories observing lines from neutrals.

 The DKIST telescope \cite{Rast21} is now in nominal operations, and it is revealing the great potentiality of the suite of instruments.
Additionally, the coordination with Solar Orbiter is a great opportunity to be taken \cite{martinez2020}. Prominence observations at high temporal and spatial scales obtained with both observatories, will sample entirely the emissions from chromospheric  to coronal temperatures, that is, the whole prominence and prominence--corona interface.
It would hopefully be possible to investigate effects PIP in the mass cycle in threads, and provide more insights into the whole stability of the prominence.   This could be addressed by the combination of neutral the H and He absorption properties within the HRIEUV band and the He\,{\sc i} measurements from CryoNIRSP.

The Solar C mission (JAXA), to be launched in 2028, will carries  an EUV--UV spectrometer  (EUVST) with 0.4'' spatial resolution and a fraction of a second cadence. A wide wavelength range is recorded by four spectral bands allowing for the measure of lines profile of neutrals (e.g. H, He) and ions in the temperature range of about 0.01\,--20 MK. These measures will allow, among others, to minimize the smoothing effect of an instrumental spatial resolution  larger than the observed fine structure, such as a better determination of velocity waves' amplitude and deduction of the local magnetic field, spicules and prominences dynamics, and other plasma properties at fine scales.  Potentially, ion--neutral Doppler  shift could be determined.

Integrating EUVST data to those of DKIST through coordinated observations will result in  spectral line measures from multiple ions and neutrals, for studying the profile properties and the link between magnetic field with the ion--neutral decoupling.
Certainly, the requirement for the precision of the measures (spectral, spatial, and temporal) is very high and, for this, careful co-registration and stability of the observations are needed.

The EST telescope \cite{noda_european_2022} will have, as one of its top science goals, the understanding of the ion--neutrals coupling in PIP. The suite of instruments proposed will provide data at 0.1'' spatial resolution and few-second cadence (e.g. Ca\,{\sc ii} 854.2, Ca\,{\sc ii} K, H\,{\sc i} 397, He\,{\sc i} 1083, and H$\alpha$) suitable for the studies including the high resolution Doppler shift and width, mass loss through draining, and magnetic field vector measurements mentioned above. 

As shown here, we are now living a privileged period thanks to the richness and complement of the available data provided by the coordination of the various space and ground-based observatories. We need to take advantage of this opportunity to have the maximum of return from this data in the frame of the PIP physics. This is also a great opportunity to improve further the realism of the prominences modelling, by increasing their complexity in the ways that have been suggested here. 
These models  need to provide, as output, additional observational elements to be realistically measured using the available observations. Only in this way, can models be correctly validated. 
As we have seen, often the spatial and temporal scales of the phenomena investigated by the simulations do not reflect, or are at the limit of, those available from the observations. It is not straightforward to solve this problem, and this should become a priority to be addressed in the coming years.

\ack{This publication is part of the R+D+i project PID2020-112791GB-I00, financed by
MCIN/AEI/10.13039/501100011033.
M.\ Luna acknowledges support through the Ram\'on y Cajal fellowship RYC2018-026129-I from the Spanish Ministry of Science and Innovation, the Spanish National Research Agency (Agencia Estatal de Investigaci\'on), the European Social Fund through Operational Program FSE 2014 of Employment, Education and Training and the Universitat de les Illes Balears. 
%
%
%
This research was supported by the International Space Science Institute
(ISSI) in Bern, through ISSI International Team project 457 (The Role of Partial Ionization in the Formation,
Dynamics and Stability of Solar Prominences).}


\bibliographystyle{RS}
\bibliography{pip.bib}

\end{document}